\documentclass[USenglish]{article}

\ifx\directlua\undefined\ifx\XeTeXcharclass\undefined
  \else\RequirePackage[no-math]{fontspec}[2017/03/31]\fi 
  \else\RequirePackage[no-math]{fontspec}[2017/03/31]\fi 
\usepackage[big]{dgruyter}
\usepackage{xcolor}
\usepackage{import}
\usepackage{subcaption}
\usepackage{multirow}
\usepackage{makecell}

\baretabulars

\begin{document}

  \articletype{Research Article}

  \author*[1]{Julian Schneider}
  \author[2]{Balint Varga}
  \author[3]{Sören Hohmann} 
  \runningauthor{Julian Schneider et al.}
  \affil[1]{Institute of Control Systems (IRS), Karlsruhe Institute of Technology (KIT), e-mail: julian.schneider@kit.edu}
  \title{Cooperative Trajectory Planning: Principles for Human-Machine System Design on Trajectory Level}
  \runningtitle{Cooperative trajectory planning}
  \abstract{This paper explores cooperative trajectory planning approaches within the context of human-machine shared control. In shared control research, it is typically assumed that the human and the automation use the same reference trajectory to stabilize the coupled system. However, this assumption is often incorrect, as they usually follow different trajectories, causing control conflicts at the action level that have not been widely researched. To address this, it is logical to extend shared control concepts to include human-machine interaction at the trajectory-level before action execution, resulting in a unified reference trajectory for both human and automation. This paper begins with a literature overview on approaches of cooperative trajectory planning. It then presents an approach of finding a joint trajectory by modelling cooperative trajectory planning as an agreement process. A generally valid system structure is proposed for this purpose. Finally, it proposes concepts to implement cooperative trajectory planning as an agreement process.}
  \received{June 20, 2024}
  \accepted{October 14, 2024}
  \journalname{at - Automatisierungstechnik}
  \journalyear{2024}
  \journalvolume{..}
  \journalissue{Special issue on human-in-the-loop systems}
  \startpage{1}
  \aop

  %
  \contributioncopyright[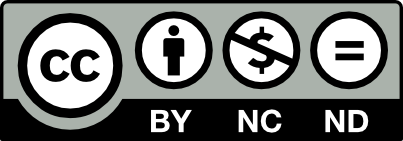]{2024}{\text{This is an Accepted Manuscript of an article submitted in at-Automatisierungstechnik journal published by De Gruyter.}}{\text{License: CC BY-NC-ND}}

\maketitle

\section{Introduction} 
Through the synergistic use of human capabilities together with automation, interactive human-machine systems promise performance gains and a reduction in human workload, leading to a symbiotic state where human cooperation seamlessly integrates with automation operation. \cite{ingaHumanMachineSymbiosisMultivariate2022}.

\begin{figure}
  \centering
  \begin{subfigure}[b]{0.3\textwidth}
      \centering
      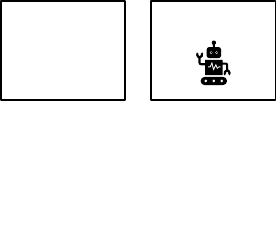
      \caption{General shared control system structure.}
      \label{fig:shared-control}
  \end{subfigure}
  \hfill
  \begin{subfigure}[b]{0.3\textwidth}
      \centering
      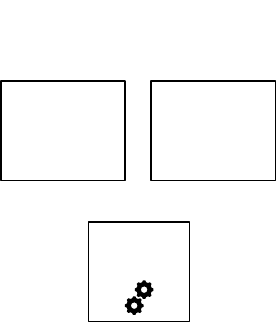
      \caption{General shared control system structure with common reference trajectory.}
      \label{fig:shared-control-reference}
  \end{subfigure}
  \hfill
  \begin{subfigure}[b]{0.3\textwidth}
    \centering
    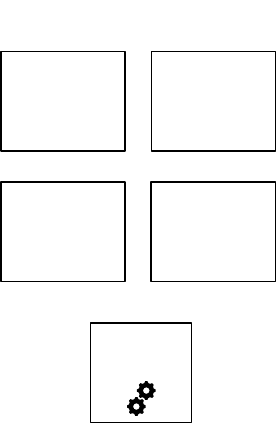
    \caption{Shared control control structure supplemented by an overlying trajectory level.}
    \label{fig:shared-control-trajectory-level}
  \end{subfigure}
  \caption{System structure of shared control without a common reference trajectory (left), with a common, given reference trajectory (middle) and with an additional trajectory level to find a common reference trajectory jointly (right).}
     \label{fig:shared-control-variants}
\end{figure}

Such interactive human-machine systems have been studied and developed in the past in the context of driver assistance systems \cite{marcanoReviewSharedControl2020} and teleoperation systems \cite{li_ClassificationNew_2023} under the term \textit{shared control}. Interactive wheelchair control is another frequently investigated application in this area of research \cite{philips_AdaptiveShared_2007, carlson_HumanwheelchairCollaboration_2008,kucukyilmaz_LearningShared_2018}. In a shared control system, a human and an automation can influence a system with their control input and both agents receive feedback from the system via output variables \cite{flad_SteeringDriver_2014}. This general structure is shown in Fig.~\ref{fig:shared-control}.

A necessary condition for the shared control system structure shown in Fig.~\ref{fig:shared-control} is that the human and the automation require the same reference trajectory as information (see Fig.~\ref{fig:shared-control-reference}). In most studies it is assumed that the reference trajectory is available to both agents. The reference trajectory generally comes from the environment, for example, the lane center in assisted driving \cite{claussmann_ReviewMotion_2020}. However, research on human driving behavior indicates that not all drivers consistently choose the center of the lane as their reference trajectory; rather, their paths are spread throughout the entire lane \cite{gunayCarFollowingTheory2007,benloucifCooperativeTrajectoryPlanning2019}. In addition, Delpiano et al. describe the so-called \textit{collateral anomaly}, according to which drivers behave differently in the lateral position of the lane when there is a vehicle directly next to them compared to no vehicle in the adjacent lane \cite{delpiano_CharacteristicsLateral_2015}. Furthermore, there are shared control systems that operate in an unstructured environment in which, for example, references do not exist \cite{vargaValidationCooperativeSharedControl2020,vargaLimitedInformationShared2023a}. There, a joint trajectory must first be found. Consequently, in experiments with deviating reference trajectories between the human and the automation, control conflicts between the two agents are reported: “subjects (...) fought the guiding forces (of the automation) indicating the automation trajectories did not optimally match the driver's trajectories.” \cite[p. 24]{abbinkHapticSharedControl2012}. It therefore requires an additional level above the action level in which a joint trajectory must be determined (see Fig.~\ref{fig:shared-control-trajectory-level}). The interaction between the human and the automation is therefore extended to the trajectory level. This doesn't mean that there is no interaction on the execution level like in shared control systems any more. In the eyes of the authors the interaction on execution level is still necessary in safety assistance systems or shared control applications where the workload in the form of control input is shifted between the human and the automation. The extension of the interaction to the trajectory level is rather necessary for applications where no common trajectory for both the human and the automation is available but first has to be jointly determined. This paper focuses on the extension of the interaction to the trajectory level and only considers the interaction at this level for simplification. However, this is not intended to exclude applications in which interaction takes place at both the trajectory level and the execution level. A possible application in which interaction takes place on both level could be, for example, the considered application in \cite{varga_ControlLarge_2019} in the context of road maintenance: at the trajectory level, an agreement on a common trajectory is first found, which is then controlled cooperatively by human and automation using the shared control methods presented therein.

The human-machine systems community is looking for generalizable forms of description for human-machine systems in the form of level models \cite{abbink_TopologyShared_2018, rothfussConceptHumanMachineNegotiation2019,flemisch_JoiningBlunt_2019}. A trajectory level for agreeing on a joint trajectory is explicitly considered here. However, the level models are generic in nature and do not include a procedure for agreeing on a joint trajectory.

In order to find a joint trajectory, the literature attempts to use intention recognition methods to recognize human behavior as a reference and to control at it \cite{corteville_HumaninspiredRobot_2007, erden_HumanIntentDetection_2010, li_HumanRobot_2014,varga_ControlLarge_2019}.

Within the shared control research field, only a few studies to date have designed cooperative trajectory planning in which an interaction between the human and the automation takes place at the trajectory level itself. From the authors' point of view, this includes the consideration of the human's and the automation's desired trajectories as input variables as well as a fusion of both desired trajectories as an output variable. Interaction within the trajectory level itself is particularly necessary if no trajectory information is available from the environment \cite{vargaValidationCooperativeSharedControl2020,vargaLimitedInformationShared2023a,schneider_NegotiationbasedCooperative_2022} and if both agents have an equal influence on the choice of trajectory, e.g. due to complementary environmental perception. To avoid control conflicts due to different reference trajectories, it is therefore necessary that there is an agreement on the joint trajectory, i.e. a kind of guarantee that both agents follow the same reference trajectory.

In this paper, approaches from the literature on cooperative trajectory planning will be presented and compared on the basis of the way in which the trajectory requests of the human and the automation are taken into account (Section \ref{sec:related-work}). The comparison results in a narrowing down of the concept of cooperative trajectory planning towards finding an agreement on a joint trajectory. Section \ref{sec:requirements} presents design requirements for this. Section \ref{sec:system-structure} presents a system design structure for reaching an agreement on a joint trajectory and Section \ref{sec:methods} deals with solution concepts of finding such an agreement.

\section{Literature overview on cooperative trajectory planning} \label{sec:related-work}
Many works can be found in the literature in the context of "human-machine interactive trajectory planning", "cooperative" or "collaborative trajectory planning". In the broadest sense, all these trajectory planning approaches work "cooperatively" according to the definition of \textit{cooperation} according to Hoc \cite{hoc_CognitiveApproach_2001}. The question is, however, how does the type of cooperation at trajectory level look in detail in each case? Are the trajectory requests of both agents taken into account or only those of one of them? If both are taken into account: How are the trajectory requests combined? The following question then arises: Is the determined trajectory for the overall system executed jointly by both agents or only by one agent alone?

\begin{figure}
  \centering
  \begin{subfigure}[b]{0.3\textwidth}
      \centering
\begingroup%
  \makeatletter%
  \providecommand\color[2][]{%
    \errmessage{(Inkscape) Color is used for the text in Inkscape, but the package 'color.sty' is not loaded}%
    \renewcommand\color[2][]{}%
  }%
  \providecommand\transparent[1]{%
    \errmessage{(Inkscape) Transparency is used (non-zero) for the text in Inkscape, but the package 'transparent.sty' is not loaded}%
    \renewcommand\transparent[1]{}%
  }%
  \providecommand\rotatebox[2]{#2}%
  \newcommand*\fsize{\dimexpr\f@size pt\relax}%
  \newcommand*\lineheight[1]{\fontsize{\fsize}{#1\fsize}\selectfont}%
  \ifx\svgwidth\undefined%
    \setlength{\unitlength}{77.17890048bp}%
    \ifx\svgscale\undefined%
      \relax%
    \else%
      \setlength{\unitlength}{\unitlength * \real{\svgscale}}%
    \fi%
  \else%
    \setlength{\unitlength}{\svgwidth}%
  \fi%
  \global\let\svgwidth\undefined%
  \global\let\svgscale\undefined%
  \makeatother%
  \begin{picture}(1,1.45765228)%
    \lineheight{1}%
    \setlength\tabcolsep{0pt}%
    \put(0,0){\includegraphics[width=\unitlength,page=1]{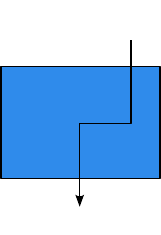}}%
    \put(0.81096598,1.2834625){\color[rgb]{0,0,0}\makebox(0,0)[t]{\lineheight{1.25}\smash{\begin{tabular}[t]{c}$\widehat{\mathfrak{T}}_\text{H}$\end{tabular}}}}%
    \put(0.4858841,0.03887076){\color[rgb]{0,0,0}\makebox(0,0)[t]{\lineheight{1.25}\smash{\begin{tabular}[t]{c}$\mathfrak{T}_\text{A}$\end{tabular}}}}%
  \end{picture}%
\endgroup%

      \caption{Leader-follower structure}
      \label{fig:boink}
  \end{subfigure}
  \begin{subfigure}[b]{0.3\textwidth}
    \centering
\begingroup%
  \makeatletter%
  \providecommand\color[2][]{%
    \errmessage{(Inkscape) Color is used for the text in Inkscape, but the package 'color.sty' is not loaded}%
    \renewcommand\color[2][]{}%
  }%
  \providecommand\transparent[1]{%
    \errmessage{(Inkscape) Transparency is used (non-zero) for the text in Inkscape, but the package 'transparent.sty' is not loaded}%
    \renewcommand\transparent[1]{}%
  }%
  \providecommand\rotatebox[2]{#2}%
  \newcommand*\fsize{\dimexpr\f@size pt\relax}%
  \newcommand*\lineheight[1]{\fontsize{\fsize}{#1\fsize}\selectfont}%
  \ifx\svgwidth\undefined%
    \setlength{\unitlength}{77.17890048bp}%
    \ifx\svgscale\undefined%
      \relax%
    \else%
      \setlength{\unitlength}{\unitlength * \real{\svgscale}}%
    \fi%
  \else%
    \setlength{\unitlength}{\svgwidth}%
  \fi%
  \global\let\svgwidth\undefined%
  \global\let\svgscale\undefined%
  \makeatother%
  \begin{picture}(1,1.45765228)%
    \lineheight{1}%
    \setlength\tabcolsep{0pt}%
    \put(0,0){\includegraphics[width=\unitlength,page=1]{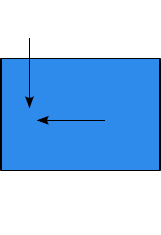}}%
    \put(0.63083469,0.75797795){\color[rgb]{0,0,0}\makebox(0,0)[t]{\lineheight{1.25}\smash{\begin{tabular}[t]{c}\footnotesize \texttt{Occurence\_Auto}\end{tabular}}}}%
    \put(0.18313438,1.29080036){\color[rgb]{0,0,0}\makebox(0,0)[t]{\lineheight{1.25}\smash{\begin{tabular}[t]{c}$\mathfrak{T}_\text{A}^*$\end{tabular}}}}%
    \put(0.4930693,0.04858663){\color[rgb]{0,0,0}\makebox(0,0)[t]{\lineheight{1.25}\smash{\begin{tabular}[t]{c}$\mathfrak{T}_\text{A}$\end{tabular}}}}%
    \put(0,0){\includegraphics[width=\unitlength,page=2]{fig5_huang.pdf}}%
  \end{picture}%
\endgroup%

    \caption{Superimposed structure}
    \label{fig:huang}
  \end{subfigure}
  \begin{subfigure}[b]{0.3\textwidth}
      \centering
\begingroup%
  \makeatletter%
  \providecommand\color[2][]{%
    \errmessage{(Inkscape) Color is used for the text in Inkscape, but the package 'color.sty' is not loaded}%
    \renewcommand\color[2][]{}%
  }%
  \providecommand\transparent[1]{%
    \errmessage{(Inkscape) Transparency is used (non-zero) for the text in Inkscape, but the package 'transparent.sty' is not loaded}%
    \renewcommand\transparent[1]{}%
  }%
  \providecommand\rotatebox[2]{#2}%
  \newcommand*\fsize{\dimexpr\f@size pt\relax}%
  \newcommand*\lineheight[1]{\fontsize{\fsize}{#1\fsize}\selectfont}%
  \ifx\svgwidth\undefined%
    \setlength{\unitlength}{86.5320219bp}%
    \ifx\svgscale\undefined%
      \relax%
    \else%
      \setlength{\unitlength}{\unitlength * \real{\svgscale}}%
    \fi%
  \else%
    \setlength{\unitlength}{\svgwidth}%
  \fi%
  \global\let\svgwidth\undefined%
  \global\let\svgscale\undefined%
  \makeatother%
  \begin{picture}(1,1.30009674)%
    \lineheight{1}%
    \setlength\tabcolsep{0pt}%
    \put(0,0){\includegraphics[width=\unitlength,page=1]{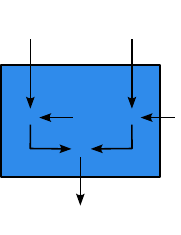}}%
    \put(0.73056226,1.15128049){\color[rgb]{0,0,0}\makebox(0,0)[t]{\lineheight{1.25}\smash{\begin{tabular}[t]{c}$\widehat{\mathfrak{T}}_\text{H}$\end{tabular}}}}%
    \put(0.3417019,0.69126272){\color[rgb]{0,0,0}\makebox(0,0)[t]{\lineheight{1.25}\smash{\begin{tabular}[t]{c}$1-\sigma$\end{tabular}}}}%
    \put(0.94841367,0.69126272){\color[rgb]{0,0,0}\makebox(0,0)[t]{\lineheight{1.25}\smash{\begin{tabular}[t]{c}$\sigma$\end{tabular}}}}%
    \put(0.16835512,1.15128049){\color[rgb]{0,0,0}\makebox(0,0)[t]{\lineheight{1.25}\smash{\begin{tabular}[t]{c}$\mathfrak{T}_\text{A}^*$\end{tabular}}}}%
    \put(0,0){\includegraphics[width=\unitlength,page=2]{fig5_benloucif.pdf}}%
    \put(0.42721718,0.04334147){\color[rgb]{0,0,0}\makebox(0,0)[t]{\lineheight{1.25}\smash{\begin{tabular}[t]{c}$\mathfrak{T}_\text{A}$\end{tabular}}}}%
    \put(0,0){\includegraphics[width=\unitlength,page=3]{fig5_benloucif.pdf}}%
  \end{picture}%
\endgroup%

      \caption{Additive controlled structure}
      \label{fig:benloucif}
  \end{subfigure}
  \begin{subfigure}[b]{0.3\textwidth}
    \centering
\begingroup%
  \makeatletter%
  \providecommand\color[2][]{%
    \errmessage{(Inkscape) Color is used for the text in Inkscape, but the package 'color.sty' is not loaded}%
    \renewcommand\color[2][]{}%
  }%
  \providecommand\transparent[1]{%
    \errmessage{(Inkscape) Transparency is used (non-zero) for the text in Inkscape, but the package 'transparent.sty' is not loaded}%
    \renewcommand\transparent[1]{}%
  }%
  \providecommand\rotatebox[2]{#2}%
  \newcommand*\fsize{\dimexpr\f@size pt\relax}%
  \newcommand*\lineheight[1]{\fontsize{\fsize}{#1\fsize}\selectfont}%
  \ifx\svgwidth\undefined%
    \setlength{\unitlength}{84.88902797bp}%
    \ifx\svgscale\undefined%
      \relax%
    \else%
      \setlength{\unitlength}{\unitlength * \real{\svgscale}}%
    \fi%
  \else%
    \setlength{\unitlength}{\svgwidth}%
  \fi%
  \global\let\svgwidth\undefined%
  \global\let\svgscale\undefined%
  \makeatother%
  \begin{picture}(1,1.32525961)%
    \lineheight{1}%
    \setlength\tabcolsep{0pt}%
    \put(0,0){\includegraphics[width=\unitlength,page=1]{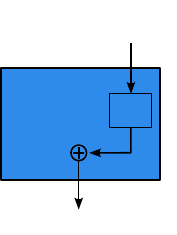}}%
    \put(0.73799646,1.14592135){\color[rgb]{0,0,0}\makebox(0,0)[t]{\lineheight{1.25}\smash{\begin{tabular}[t]{c}$\Delta\widehat{\mathfrak{T}}_\text{H}$\end{tabular}}}}%
    \put(0.44430453,0.03750527){\color[rgb]{0,0,0}\makebox(0,0)[t]{\lineheight{1.25}\smash{\begin{tabular}[t]{c}$\mathfrak{T}_\text{A}$\end{tabular}}}}%
    \put(0.16491554,1.14592135){\color[rgb]{0,0,0}\makebox(0,0)[t]{\lineheight{1.25}\smash{\begin{tabular}[t]{c}$\mathfrak{T}_\text{A}^*$\end{tabular}}}}%
    \put(0,0){\includegraphics[width=\unitlength,page=2]{fig5_losey.pdf}}%
    \put(0.73899463,0.67088618){\color[rgb]{0,0,0}\makebox(0,0)[t]{\lineheight{1.25}\smash{\begin{tabular}[t]{c}$\mu$\end{tabular}}}}%
  \end{picture}%
\endgroup%

    \caption{Additive deforming structure}
    \label{fig:losey}
  \end{subfigure}
  \begin{subfigure}[b]{0.3\textwidth}
    \centering
\begingroup%
  \makeatletter%
  \providecommand\color[2][]{%
    \errmessage{(Inkscape) Color is used for the text in Inkscape, but the package 'color.sty' is not loaded}%
    \renewcommand\color[2][]{}%
  }%
  \providecommand\transparent[1]{%
    \errmessage{(Inkscape) Transparency is used (non-zero) for the text in Inkscape, but the package 'transparent.sty' is not loaded}%
    \renewcommand\transparent[1]{}%
  }%
  \providecommand\rotatebox[2]{#2}%
  \newcommand*\fsize{\dimexpr\f@size pt\relax}%
  \newcommand*\lineheight[1]{\fontsize{\fsize}{#1\fsize}\selectfont}%
  \ifx\svgwidth\undefined%
    \setlength{\unitlength}{81.1125275bp}%
    \ifx\svgscale\undefined%
      \relax%
    \else%
      \setlength{\unitlength}{\unitlength * \real{\svgscale}}%
    \fi%
  \else%
    \setlength{\unitlength}{\svgwidth}%
  \fi%
  \global\let\svgwidth\undefined%
  \global\let\svgscale\undefined%
  \makeatother%
  \begin{picture}(1,1.38696208)%
    \lineheight{1}%
    \setlength\tabcolsep{0pt}%
    \put(0,0){\includegraphics[width=\unitlength,page=1]{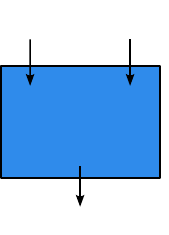}}%
    \put(0.77410084,1.22348548){\color[rgb]{0,0,0}\makebox(0,0)[t]{\lineheight{1.25}\smash{\begin{tabular}[t]{c}$\widehat{\mathfrak{T}}_\text{H}$\end{tabular}}}}%
    \put(0.45172371,0.03926747){\color[rgb]{0,0,0}\makebox(0,0)[t]{\lineheight{1.25}\smash{\begin{tabular}[t]{c}$\mathfrak{T}_\text{A}$\end{tabular}}}}%
    \put(0.17782596,1.22348548){\color[rgb]{0,0,0}\makebox(0,0)[t]{\lineheight{1.25}\smash{\begin{tabular}[t]{c}$\mathfrak{T}_\text{A}^*$\end{tabular}}}}%
    \put(0,0){\includegraphics[width=\unitlength,page=2]{fig5_schneider.pdf}}%
    \put(0.48333425,0.58602475){\color[rgb]{0,0,0}\makebox(0,0)[t]{\lineheight{1.25}\smash{\begin{tabular}[t]{c}finding\\agreement\end{tabular}}}}%
  \end{picture}%
\endgroup%

    \caption{Agreement structure}
    \label{fig:schneider}
  \end{subfigure}
     \caption{Comparison of five different instantiations of fusions of trajectory requests.}
     \label{fig:trajectory-combinations}
\end{figure}

A large group of studies attempts to use intention recognition methods to estimate the human's trajectory request and then regulate it by the automation \cite{corteville_HumaninspiredRobot_2007, erden_HumanIntentDetection_2010, li_HumanRobot_2014, gnatzig_HumanmachineInteraction_2012, boinkUnderstandingReducingConflicts2014}. There are also differences within this group in that the subsequent control of the trajectory desired by the human at the action level (Fig.~\ref{fig:shared-control-trajectory-level}) is either executed by the automation alone \cite{gnatzig_HumanmachineInteraction_2012} (the human is completely excluded from the action level here) or by the human and the automation together in the form of haptic shared control \cite{boinkUnderstandingReducingConflicts2014}. At the trajectory level, these approaches only consider the human's trajectory request, whereby the human is the leader and the automation is the follower. Fig.~\ref{fig:boink} shows this type of fusion of the human's trajectory desire with that of the automation in the form of a block diagram with an input and an output signal. Since only the estimated trajectory request $\widehat{\mathfrak{T}}_\text{H}$ of the human is considered as the input variable here (the roof denotes an estimated variable), there is no fusion of trajectory requests of the human and the automation needed and the automation considers the request of the human $\hat{\mathfrak{T}}_\text{H}$ as its trajectory  $\mathfrak{T}_\text{A}$ to execute.

However, it cannot always be guaranteed that the human will choose a safe trajectory. Huang et al. therefore present an approach in the context of assisted driving in which the automation plans its own reference and applies corrective control input to the vehicle if it detects that the vehicle is leaving the safe space defined in the approach \cite{huangHumanMachineCooperativeTrajectory2022}. The trajectory request of the automation is superimposed on the movement of the human. Fig.~\ref{fig:huang} shows this relationship as a block diagram which, in the case of a detected dangerous driving style (Occurence\_Auto=1), superimposes the corrective control input on the basis of the trajectory request $\mathfrak{T}_\text{A}$ of the automation. The trajectory request $\widehat{\mathfrak{T}}_\text{H}$ of the human is not explicitly taken into account in this approach. A similar intervention in the detection of an unsafe driving style takes place in \cite{jiang_HumanMachineCooperative_2021}. Instead of applying a purely corrective control input, an optimization-based trajectory that comes as close as possible to the human's desired trajectory is applied here.

Benloucif et al. present an approach - also in the context of assisted driving - in which control conflicts between humans and automation due to different reference trajectories are to be resolved by estimating the lateral position on the lane $y_\text{des}$ desired by the human \cite{benloucifCooperativeTrajectoryPlanning2019}. This value is then included in the trajectory planning of the automation. Furthermore, the driver's attention level $\sigma$ is determined as a driver state parameter. Depending on this attention level $\sigma$, the human's lateral position request $y_\text{des}$ is taken into account to a greater or lesser extent. Fig.~\ref{fig:benloucif} shows the procedure as a block diagram with the two trajectory desires as input variables. The fusion of the estimated trajectory desire $\widehat{\mathfrak{T}}_\text{H}$ of the human with the trajectory desire $\mathfrak{T}^*_\text{A}$ of the automation (the star denotes the optimal trajectory for the automation based on an underlying cost function) is done additively using the weighting of the driver state parameter $\sigma$. One disadvantage of this linear blending approach is the fact that the result of linearly combining two safe trajectories can result in an unsafe trajectory \cite{trautman_AssistivePlanning_2015}.

In \cite{losey_TrajectoryDeformations_2018}, a physical human-robot interaction (pHRI) application is considered in which a human operates a haptic device in order to control a robot together with an automation. The system initially follows the target trajectory of the automation. The human can cause a trajectory deformation by exerting a force. This force is measured by the automation and an optimization-based approach with a minimum-jerk model is used to calculate a trajectory deformation corresponding to the human's request. The fixed parameter $\mu$ determines the amplitude-like expansion of the human's trajectory deformation. The higher $\mu$, the greater the consideration of the person's desire for change and the greater the deformation. The parameter $\mu$ can be regarded as a kind of amplification, even if the relationship between trajectory deformation and the parameter $\mu$ is nonlinear \cite{losey_TrajectoryDeformations_2018}. Fig.~\ref{fig:losey} shows the corresponding block diagram represented by $\Delta \widehat{\mathfrak{T}}_\text{H}$ and the fixed parameter $\mu$ as gain for the consideration of the human's desire to change motion. The disadvantage of the linear blending is again the potential unsafe trajectory by means of the linear combination of two safe trajectories \cite{trautman_AssistivePlanning_2015}.

In contrast to \cite{boinkUnderstandingReducingConflicts2014}, \cite{benloucifCooperativeTrajectoryPlanning2019} and \cite{huangHumanMachineCooperativeTrajectory2022} have in common that the automation plans its own reference trajectory and forms a compromise with the human reference trajectory. What \cite{benloucifCooperativeTrajectoryPlanning2019} and \cite{huangHumanMachineCooperativeTrajectory2022} also have in common is that automation uses safety considerations to decide which agent has how much influence on the choice of trajectory. In \cite{losey_TrajectoryDeformations_2018}, the degree of human influence is determined using the fixed parameter $\mu$. In \cite{schneider_NegotiationbasedCooperative_2022}, on the other hand, the influence of both agents is neither fixed from the outset nor determined by the automation, but continuously negotiated by both agents. The current trajectory request of the human is estimated at each point in time and combined with the current trajectory request of the automation using negotiation theory. The resulting trajectory for the overall system depends on the compliance of the two agents. This approach is particularly advantageous when applications are less time- and safety-critical, such as agreeing on a joint trajectory when a robot accompanies a patient \cite{schneider_NegotiationbasedCooperative_2022}. Fig.~\ref{fig:schneider} shows the corresponding block diagram. The special feature of this approach is that an agreement is reached on the joint trajectory between the human and the automation. In contrast to the leader-follower principle, this corresponds to an emancipated approach. A characteristic of this approach is that both agents are given the opportunity to react to the other agent's desire to move. The result is an agreement on a joint trajectory after the reactions have taken place.

The comparison of the block diagrams of the trajectory request fusions shown in Fig.~\ref{fig:trajectory-combinations} shows on the one hand that the approaches \cite{boinkUnderstandingReducingConflicts2014} and \cite{huangHumanMachineCooperativeTrajectory2022} only consider the trajectory request of one agent at the trajectory level. In addition, it can be seen that in the approaches \cite{benloucifCooperativeTrajectoryPlanning2019} and \cite{losey_TrajectoryDeformations_2018} the fusion of the trajectory requests is determined by the automation. In these cases, the human has no say in the way in which the two trajectory requests are fusioned or cannot provide feedback on the fusion method. Such a fusion that is determined by one agent is advantageous in safety- and time-critical tasks, as in these cases there is no time to negotiate. Here, however, it must always be ensured that the fusion is correctly and in a way that the human accepts or finds helpful.

Compared to the literature, the approach in \cite{schneider_NegotiationbasedCooperative_2022} represents a new approach to cooperation at trajectory level. The core idea of this approach is the consideration of the human's and the automation's trajectory request as an emancipated agreement on a joint trajectory. For the further development of such an emancipated agreement on a joint trajectory, general requirements will be defined in the following section and possible methods for the implementation of cooperative trajectory finding that implement an agreement process will be presented in Section \ref{sec:methods}.

\section{Requirements for the design of cooperative trajectory planning} \label{sec:requirements}

In shared control applications in which the reference trajectory is not derived from environmental information \cite{vargaValidationCooperativeSharedControl2020,vargaLimitedInformationShared2023a} and in which humans and machines have equal influence on the choice of trajectory, e.g. due to complementary environmental perception, modeling trajectory planning as a cooperative agreement process already ensures in the design that both agents choose the same trajectory as a joint reference trajectory. This in turn avoids control conflicts between humans and automation. As an objection, one could argue that this only moves the conflict between human and automation one level up to the trajectory level. This is true, but the interaction on trajectory level provides a longer time horizon for agreement in contrast to the execution level. In addition - and this is the more important property of the extension of the interaction to the trajectory level - agreements at the execution level can lead to situations in which the human and the automaiton are in a kind of tug-of-war according to the Nash equilibrium: from the outward perspective, the coupled system appears to be moving in a stationary state on an agreed trajectory. However, in the inside the automation and the human apply a control input to the system that is equal in magnitude but different in sign. Although this keeps the system in a stationary state on the output variables, inside the system the automation and the human apply a control input that is greater zero. These control conflicts can be avoided by agreeing on a common trajectory by means of interaction on the trajectory level. Therefore, the first requirement proposed is the modeling of cooperative trajectory finding as an agreement process.

Secondly, this agreement process should take place on an equal footing, i.e. the human and the automation have equal influence within this agreement process. Neither agent may outvote the other in any way. In contrast to leader-follower approaches from the literature, this is therefore an emancipated approach.

Thirdly, it is not a safety- and time-critical application in which a decision may have to be made within a short period of time, as it cannot be guaranteed that an agreement will be reached within the required time. Possible applications in this context can be the considered application of road maintenance in \cite{varga_ControlLarge_2019} that could be extended with an additional interaction on the trajectory level where first a common trajectory could be found. Another application could be the cooperative trajectory planning presented in \cite{schneider_NegotiationbasedCooperative_2022}.

Fourthly, the procedure for cooperative trajectory finding in the form of an agreement process should be as generalizable and application-independent as possible. 

The four requirements for cooperative trajectory finding can be summarized as follows:
\begin{enumerate}
  \item Modeling as an agreement process.
  \item Emancipated approach (vs. leader-follower): Both agents have equal influence or are equal (humans must not be outvoted by automation in the end).
  \item No safety and time-critical application.
  \item Generalizability, i.e. independence from specific application.
\end{enumerate}
The four requirements mentioned have the following implications for the design of a cooperative trajectory planning:
\begin{enumerate}
  \item As a result of the modeling as an agreement process, communication between the two agents is required. This takes place via the exchange of the system-theoretical variables information and/or energy in the form of physical variables (e.g. forces). Information exchange can also take place via spoken dialog.
  \item As a consequence of the aforementioned communication aspect, the structure requires a module that forms the automation's desire to move and a module for understanding or interpretation of the human's desire to move.
  \item As a further communication aspect, a reaction to the agent's reaction to its own trajectory request takes place during the agreement process. This reaction cycle converges until agreement is reached on the joint reference trajectory.
\end{enumerate}

\section{System structure for cooperative trajectory planning}\label{sec:system-structure}
\begin{figure}[t]
  \centering
  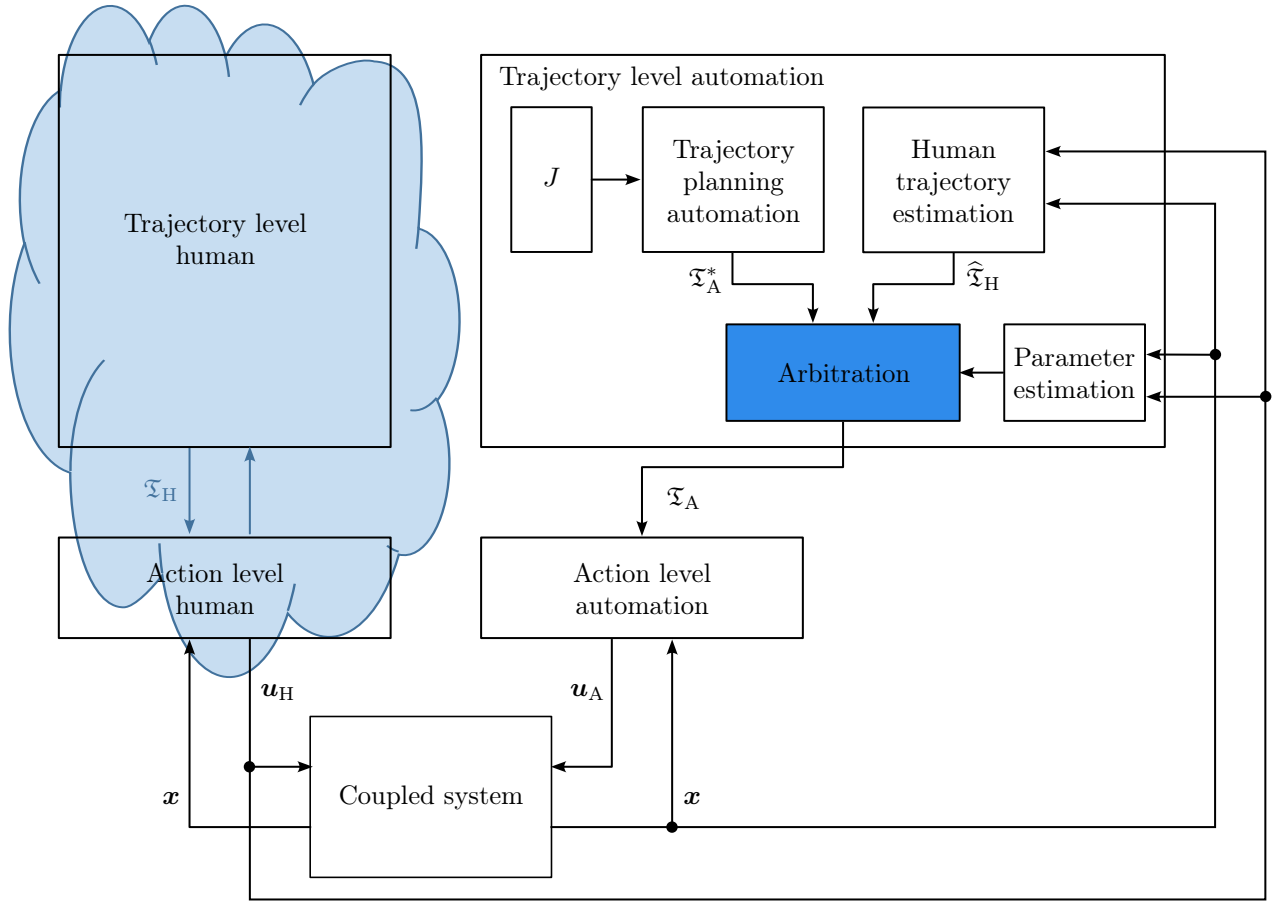
  \caption{System structure for cooperative trajectory planning at the trajectory level.}
  \label{fig:system-structure}
\end{figure}

A system structure that describes cooperative trajectory planning at the trajectory level can be derived from the approaches presented in Section \ref{sec:related-work}, which takes into account the trajectory planning requirements of the human and the automation from Section \ref{sec:requirements} as well as the mentioned implications. The proposed system structure is shown in Fig.~\ref{fig:system-structure} and describes not only the trajectory planning approach by modeling it as an agreement process but both the approaches \cite{benloucifCooperativeTrajectoryPlanning2019, losey_TrajectoryDeformations_2018} and \cite{schneider_NegotiationbasedCooperative_2022}. Only the specialization of the fusion or arbitration block (blue) differentiates between the individual approaches.

The trajectory levels in Fig.~\ref{fig:system-structure} describe the trajectory planning of the two agents and the action levels describe the execution of the planned trajectory. This is achieved by the input of control variables $\boldsymbol{u}_\text{H}$ and $\boldsymbol{u}_\text{A}$ to the coupled system, from which both agents receive the system state $\boldsymbol{x}$ as feedback. The structure for cooperative trajectory planning presented in this paper has one important characteristic: Humans typically do not employ an explicit reference trajectory for their motion planning \cite{todorovOptimalFeedbackControl2002}. Therefore, the human’s trajectory level and action level are drawn in a cloud. However, adopting collaborative trajectory planning enhances the development of a more human-centric automation.

Cooperative trajectory planning can then on automation side be further divided into a trajectory planning subsystem for automation with a cost function $J$ as input, a subsystem for trajectory estimation for the human's desired trajectory and an arbitration module for merging the two desired trajectories. In addition, the arbitration module can optionally be preceded by a subsystem for determining parameters (e.g. human attention) that are used for arbitration. Cooperative trajectory planning is then carried out using the following four steps:
\begin{enumerate}
  \item \textbf{Choice of a trajectory planning method:} The first step is to select a trajectory planner that fulfills the application requirements e.g., computational effort, real-time capability or (non-) parametric method. Suitable planners can be found in \cite[Section 1.B]{benloucifCooperativeTrajectoryPlanning2019}. Each trajectory planning method has a set of parameters that are derived from current state variables (e.g., position, speed), setting parameters (e.g., prediction horizon) and desired parameters (e.g., execution time, maximum longitudinal or lateral acceleration), which can also be specified via parameters in a cost function $J$ (e.g., penalization of the execution time for the shortest possible execution time, analogous formulation for jerk, lateral acceleration, etc.).
  \item \textbf{Subsystem \textit{trajectory planning automation}:} In this subsystem, the automation calculates a trajectory for the set time horizon using the method selected in step 1 via a cost function $J$. To achieve this, the parameter values required to meet the automation's criteria are established. The desired trajectory of the automation is denoted by $\mathfrak{T}^*_\text{A}$. The asterisk indicates the optimal trajectory calculated for the automation.
  \item \textbf{Subsystem \textit{human trajectory estimation}:} The presented works \cite{benloucifCooperativeTrajectoryPlanning2019,losey_TrajectoryDeformations_2018,schneider_NegotiationbasedCooperative_2022} that consider the human's desire to move include an application-specific estimation module for estimating the human's desire to move. In this step, the selected trajectory planning method from step 1 determines the set of parameters that must be estimated from the measured human behavior that is mainly described by the human control variables and the system state. The estimation module estimates the parameters either offline from previous recorded data \cite{boinkUnderstandingReducingConflicts2014} or online \cite{benloucifCooperativeTrajectoryPlanning2019}. The estimated trajectory request of the human is denoted by $\widehat{\mathfrak{T}}_\text{H}$.
  \item \textbf{Subsystem \textit{arbitration}:} All the works presented that take into account the trajectory request of the human and the automation have to fuse these two requests in a certain way. This module determines how the two trajectory requests are fused or arbitrated to form a trajectory $\mathfrak{T}_\text{A}$ for the coupled system. A subsystem with a \textit{parameter estimation} in front of the fusion can be the determination of additional parameters (e.g. human state parameters), based on which the arbitration is performed. In \cite{benloucifCooperativeTrajectoryPlanning2019} it is the determination of the driver attention level $\sigma \in [0,1]$.
\end{enumerate}

If the block diagrams from Fig.~\ref{fig:benloucif}, \ref{fig:losey} or \ref{fig:schneider} are used for the arbitration block (blue) in Fig. \ref{fig:system-structure}, the different cooperative trajectory planning approaches \cite{benloucifCooperativeTrajectoryPlanning2019}, \cite{losey_TrajectoryDeformations_2018} or \cite{schneider_NegotiationbasedCooperative_2022} result. The 

The following section presents and discusses solution concepts for cooperative trajectory planning in compliance with the requirements mentioned in Section \ref{sec:requirements} especially for cooperatively finding an agreement on a joint trajectory.

\section{Solution concepts for finding a cooperative agreement on a joint trajectory} \label{sec:methods}
In this work, a cooperative trajectory planning based on the agreement on a common trajectory was presented for applications in which no reference trajectory results from the environment or other boundary conditions. As starting point there are two possible options for choosing the approach to emancipated cooperative trajectory planning, taking into account an agreement process: The knowledge of how humans and automation will agree, i.e. the consensus on a joint trajectory, is either already available or the consensus must first be determined. The first case (consensus is available) can be determined in experiments, for example, and described in a model. In the second case, without prior knowledge of the consensus, the consensus must first be determined. The case in which knowledge of the agreement is already available is used in the modeling of human-machine interaction as a differential game using the Nash equilibrium \cite{flad_SteeringDriver_2014}. The humans and the automation are modeled here as rational agents in the form of optimal controllers, and the behavior of humans and machines results from the Nash strategy. This procedure of postulating behavior in the case of consensus as a Nash strategy often takes place at the action level. The model knowledge of the behavior of humans in the case of consensus is described using cost functionals of an optimal control approach. The parameters of the cost functional performance measures are determined from data from experimental experiments via inverse optimization.

The fact that this involves the application of a consensus of an agreement process can be motivated on the one hand by the fact that there is a Nash equilibrium. On the other hand, it can be illustrated by the calculation method of the iterative best response for determining the optimal control values in the Nash equilibrium: Each agent alternately calculates his optimal control variable while keeping the agent's control variable fixed. This happens iteratively until the manipulated variables converge to the Nash strategy.

However, this approach cannot be applied at the trajectory level, since model knowledge about the trajectory planning behavior of humans would have to be known. In particular, the human’s cost functional of a trajectory optimization planner would have to be determined. According to \cite{todorovOptimalFeedbackControl2002}, however, humans do not act in a trajectory planning and trajectory action level, but do this in one step. Thus, model knowledge about the trajectory planning behavior of a human in the case of consensus with a machine cannot be determined. Therefore, consensus on a joint trajectory must be found before execution. In \cite{rothfussHumanMachineCooperativeDecision2022}, two methods for agreeing on a common driving maneuver are presented. These are a negotiation theory approach \cite[Section 3.2]{rothfussHumanMachineCooperativeDecision2022} and a game theoretic approach \cite[Section 3.3]{rothfussHumanMachineCooperativeDecision2022}. Both approaches consider the agreement on a discrete driving maneuver. In the case of cooperative trajectory finding, the two approaches must be extended from negotiation about a discrete event to continuous negotiation about trajectories or parameters describing trajectories.

\section{Conclusion}
In this paper, a literature overview of cooperative trajectory planning methods between a human and an automation within shared control was presented (Section \ref{sec:related-work}) and requirements were defined for a special form of cooperative trajectory planning in the form of an emancipated agreement on a joint trajectory (Section \ref{sec:requirements}). Section \ref{sec:system-structure} presented a system structure for designing cooperative trajectory planning and Section \ref{sec:methods} discussed solution concepts for cooperative trajectory planning.

The next steps are to implement the presented concepts (negotiation theory and game theory) for the continuous negotiation of a trajectory in an application and evaluate them in a study. Secondly, the flow of communication between the human and the automation needs to be examined more closely. 

As an outlook, in shared control applications where there is already an interaction on execution level and an interaction on trajectory level added there is probably expected an increase in the workload for the human and maybe a decrease in the situational awareness. This expecation can not be stated yet surely and has also to be examined in a further study.

\bibliographystyle{ieeetr}
\bibliography{literature}

\begin{thebibliography}{10}

\bibitem{ingaHumanMachineSymbiosisMultivariate2022}
J.~Inga, M.~Ruess, J.~Heinrich~Robens, T.~Nelius, S.~Rothfu{\ss}, S.~Kille, P.~Dahlinger, A.~Lindenmann, R.~Thomaschke, G.~Neumann, S.~Matthiesen, S.~Hohmann, and A.~Kiesel, ``{Human-Machine Symbiosis: A Multivariate Perspective for Physically Coupled Human-Machine Systems},'' {\em International Journal of Human-Computer Studies}, vol.~170, p.~102926, 2022.

\bibitem{marcanoReviewSharedControl2020}
M.~Marcano, S.~D{\'i}az, J.~P{\'e}rez, and E.~Irigoyen, ``A {{Review}} of {{Shared Control}} for {{Automated Vehicles}}: {{Theory}} and {{Applications}},'' {\em IEEE Transactions on Human-Machine Systems}, vol.~50, no.~6, pp.~475--491, 2020.

\bibitem{li_ClassificationNew_2023}
G.~Li, Q.~Li, C.~Yang, Y.~Su, Z.~Yuan, and X.~Wu, ``The {{Classification}} and {{New Trends}} of {{Shared Control Strategies}} in {{Telerobotic Systems}}: {{A Survey}},'' {\em IEEE Transactions on Haptics}, vol.~16, no.~2, pp.~118--133, 2023.

\bibitem{philips_AdaptiveShared_2007}
J.~Philips, J.~{del R. Millan}, G.~Vanacker, E.~Lew, F.~Galan, P.~W. Ferrez, H.~Van~Brussel, and M.~Nuttin, ``Adaptive {{Shared Control}} of a {{Brain-Actuated Simulated Wheelchair}},'' in {\em 2007 {{IEEE}} 10th {{International Conference}} on {{Rehabilitation Robotics}}}, pp.~408--414, 2007.

\bibitem{carlson_HumanwheelchairCollaboration_2008}
T.~Carlson and Y.~Demiris, ``Human-wheelchair collaboration through prediction of intention and adaptive assistance,'' in {\em 2008 {{IEEE International Conference}} on {{Robotics}} and {{Automation}}}, pp.~3926--3931, 2008.

\bibitem{kucukyilmaz_LearningShared_2018}
A.~Kucukyilmaz and Y.~Demiris, ``Learning {{Shared Control}} by {{Demonstration}} for {{Personalized Wheelchair Assistance}},'' {\em IEEE Transactions on Haptics}, vol.~11, no.~3, pp.~431--442, 2018.

\bibitem{flad_SteeringDriver_2014}
M.~Flad, J.~Otten, S.~Schwab, and S.~Hohmann, ``Steering driver assistance system: {{A}} systematic cooperative shared control design approach,'' in {\em 2014 {{IEEE International Conference}} on {{Systems}}, {{Man}}, and {{Cybernetics}} ({{SMC}})}, pp.~3585--3592, 2014.

\bibitem{claussmann_ReviewMotion_2020}
L.~Claussmann, M.~Revilloud, D.~Gruyer, and S.~Glaser, ``A {{Review}} of {{Motion Planning}} for {{Highway Autonomous Driving}},'' {\em IEEE Transactions on Intelligent Transportation Systems}, vol.~21, no.~5, pp.~1826--1848, 2020.

\bibitem{gunayCarFollowingTheory2007}
B.~Gunay, ``Car following theory with lateral discomfort,'' {\em Transportation Research Part B: Methodological}, vol.~41, no.~7, pp.~722--735, 2007.

\bibitem{benloucifCooperativeTrajectoryPlanning2019}
A.~Benloucif, A.-T. Nguyen, C.~Sentouh, and J.-C. Popieul, ``Cooperative {{Trajectory Planning}} for {{Haptic Shared Control Between Driver}} and {{Automation}} in {{Highway Driving}},'' {\em IEEE Transactions on Industrial Electronics}, vol.~66, no.~12, pp.~9846--9857, 2019.

\bibitem{delpiano_CharacteristicsLateral_2015}
R.~Delpiano, J.~Herrera~M., and J.~Coeymans~A., ``Characteristics of lateral vehicle interaction,'' {\em Transportmetrica A: Transport Science}, vol.~11, no.~7, pp.~636--647, 2015.

\bibitem{vargaValidationCooperativeSharedControl2020}
B.~Varga, A.~Shahirpour, Y.~Burkhardt, S.~Schwab, and S.~Hohmann, ``Validation of {{Cooperative Shared-Control Concepts}} for {{Large Vehicle-Manipulators}},'' in {\em 2020 {{IEEE Conference}} on {{Control Technology}} and {{Applications}} ({{CCTA}})}, pp.~542--548, 2020.

\bibitem{vargaLimitedInformationShared2023a}
B.~Varga, J.~Inga, and S.~Hohmann, ``Limited {{Information Shared Control}}: {{A Potential Game Approach}},'' {\em IEEE Transactions on Human-Machine Systems}, vol.~53, no.~2, pp.~282--292, 2023.

\bibitem{abbinkHapticSharedControl2012}
D.~A. Abbink, M.~Mulder, and E.~R. Boer, ``Haptic shared control: Smoothly shifting control authority?,'' {\em Cognition, Technology \& Work}, vol.~14, no.~1, pp.~19--28, 2012.

\bibitem{varga_ControlLarge_2019}
B.~Varga, A.~Shahirpour, S.~Schwab, and S.~Hohmann, ``Control of {{Large Vehicle-Manipulators}} with {{Human Operator}},'' {\em IFAC-PapersOnLine}, vol.~52, no.~30, pp.~373--378, 2019.

\bibitem{abbink_TopologyShared_2018}
D.~A. Abbink, T.~Carlson, M.~Mulder, J.~C.~F. {de Winter}, F.~Aminravan, T.~L. Gibo, and E.~R. Boer, ``A {{Topology}} of {{Shared Control Systems}}---{{Finding Common Ground}} in {{Diversity}},'' {\em IEEE Transactions on Human-Machine Systems}, vol.~48, no.~5, pp.~509--525, 2018.

\bibitem{rothfussConceptHumanMachineNegotiation2019}
S.~Rothfu{\ss}, R.~Schmidt, M.~Flad, and S.~Hohmann, ``A {{Concept}} for {{Human-Machine Negotiation}} in {{Advanced Driving Assistance Systems}},'' in {\em 2019 {{IEEE International Conference}} on {{Systems}}, {{Man}} and {{Cybernetics}} ({{SMC}})}, pp.~3116--3123, 2019.

\bibitem{flemisch_JoiningBlunt_2019}
F.~Flemisch, D.~A. Abbink, M.~Itoh, M.-P. {Pacaux-Lemoine}, and G.~We{\ss}el, ``Joining the blunt and the pointy end of the spear: Towards a common framework of joint action, human--machine cooperation, cooperative guidance and control, shared, traded and supervisory control,'' {\em Cognition, Technology \& Work}, vol.~4, no.~21, pp.~555--568, 2019.

\bibitem{corteville_HumaninspiredRobot_2007}
B.~Corteville, E.~Aertbelien, H.~Bruyninckx, J.~De~Schutter, and H.~Van~Brussel, ``Human-inspired robot assistant for fast point-to-point movements,'' in {\em Proceedings 2007 {{IEEE International Conference}} on {{Robotics}} and {{Automation}}}, pp.~3639--3644, 2007.

\bibitem{erden_HumanIntentDetection_2010}
M.~S. Erden and T.~Tomiyama, ``Human-{{Intent Detection}} and {{Physically Interactive Control}} of a {{Robot Without Force Sensors}},'' {\em IEEE Transactions on Robotics}, vol.~26, no.~2, pp.~370--382, 2010.

\bibitem{li_HumanRobot_2014}
Y.~Li and S.~S. Ge, ``Human--{{Robot Collaboration Based}} on {{Motion Intention Estimation}},'' {\em IEEE/ASME Transactions on Mechatronics}, vol.~19, no.~3, pp.~1007--1014, 2014.

\bibitem{schneider_NegotiationbasedCooperative_2022}
J.~Schneider, S.~Rothfu{\ss}, and S.~Hohmann, ``Negotiation-based cooperative planning of local trajectories,'' {\em Frontiers in Control Engineering}, vol.~3, 2022.

\bibitem{hoc_CognitiveApproach_2001}
J.-M. Hoc, ``Towards a cognitive approach to human-machine cooperation in dynamic situations,'' {\em International Journal of Human-Computer Studies}, vol.~54, no.~4, pp.~509--540, 2001.

\bibitem{gnatzig_HumanmachineInteraction_2012}
S.~Gnatzig, F.~Schuller, and M.~Lienkamp, ``Human-machine interaction as key technology for driverless driving - {{A}} trajectory-based shared autonomy control approach,'' in {\em 2012 {{IEEE RO-MAN}}: {{The}} 21st {{IEEE International Symposium}} on {{Robot}} and {{Human Interactive Communication}}}, pp.~913--918, 2012.

\bibitem{boinkUnderstandingReducingConflicts2014}
R.~Boink, M.~M. {van Paassen}, M.~Mulder, and D.~A. Abbink, ``Understanding and reducing conflicts between driver and haptic shared control,'' in {\em 2014 {{IEEE International Conference}} on {{Systems}}, {{Man}}, and {{Cybernetics}} ({{SMC}})}, pp.~1510--1515, 2014.

\bibitem{huangHumanMachineCooperativeTrajectory2022}
C.~Huang, H.~Huang, J.~Zhang, P.~Hang, Z.~Hu, and C.~Lv, ``Human-{{Machine Cooperative Trajectory Planning}} and {{Tracking}} for {{Safe Automated Driving}},'' {\em IEEE Transactions on Intelligent Transportation Systems}, vol.~23, no.~8, pp.~12050--12063, 2022.

\bibitem{jiang_HumanMachineCooperative_2021}
B.~Jiang, X.~Li, Y.~Zeng, and D.~Liu, ``Human-{{Machine Cooperative Trajectory Planning}} for {{Semi-Autonomous Driving Based}} on the {{Understanding}} of {{Behavioral Semantics}},'' {\em Electronics}, vol.~10, no.~8, p.~946, 2021.

\bibitem{trautman_AssistivePlanning_2015}
P.~Trautman, ``Assistive {{Planning}} in {{Complex}}, {{Dynamic Environments}}: {{A Probabilistic Approach}},'' in {\em 2015 {{IEEE International Conference}} on {{Systems}}, {{Man}}, and {{Cybernetics}}}, pp.~3072--3078, 2015.

\bibitem{losey_TrajectoryDeformations_2018}
D.~P. Losey and M.~K. O'Malley, ``Trajectory {{Deformations From Physical Human}}--{{Robot Interaction}},'' {\em IEEE Transactions on Robotics}, vol.~34, no.~1, pp.~126--138, 2018.

\bibitem{todorovOptimalFeedbackControl2002}
E.~Todorov and M.~I. Jordan, ``Optimal feedback control as a theory of motor coordination,'' {\em Nature Neuroscience}, vol.~5, no.~11, pp.~1226--1235, 2002.

\bibitem{rothfussHumanMachineCooperativeDecision2022}
S.~Rothfu{\ss}, {\em Human-{{Machine Cooperative Decision Making}}}.
\newblock PhD thesis, Karlsruher Institut f{\"u}r Technologie (KIT), Karlsruhe, 2022.

\end{thebibliography}
\end{document}